\documentclass[aps,prev,preprintnumbers,floatset,nofootinbib,onecolumn,11pt]{revtex4-1}
\usepackage[english]{babel}
\usepackage{bm}
\usepackage{times}
\usepackage{ulem}
\usepackage{hyperref}
\hypersetup{
    colorlinks=true,
    linkcolor=blue,
    filecolor=magenta,      
    urlcolor=blue,
    citecolor=blue,
    pdftitle={Sharelatex Example},
    pdfpagemode=FullScreen
    }
\usepackage{listings}
\usepackage{slashed}
\usepackage{color}
\usepackage{appendix}
\usepackage{mathtools}
\usepackage{epsfig}
\usepackage{dcolumn}
\usepackage{textcomp}
\usepackage{appendix} 
\usepackage{multirow}
\usepackage{graphicx}
\usepackage{soul}
\usepackage{subfigure}

\newcommand{\llg}{\ell_8}
\newcommand{\lgx}{\bar{\ell}_8}

\begin{document} 

\title{Resonant Lepton-Gluon Collisions at the Large Hadron Collider}

\author{Eduardo da Silva Almeida$^{1}$}
\email{eduardo.silva.almeida@usp.br}
\author{Alexandre Alves$^{1}$}
\email{aalves@unifesp.br}
\author{Oscar J. P.  Éboli$^{2,3}$}
\email{eboli@if.usp.br}
\author{F. S. Queiroz$^{4,5,6}$}
\email{farinaldo.queiroz@ufrn.br}

\affiliation{$^1$Departamento de F\'isica, Universidade Federal de S\~ao Paulo, UNIFESP, Diadema, S\~ao Paulo, Brazil}
\affiliation{$^2$ Instituto de F\'isica, Universidade de S\~ao Paulo,
R. do Mat\~ao 1371, 05508-090 S\~ao Paulo, Brazil}
\affiliation{$^3$ Departament de Fisica Quantica i Astrofisica and Institut de Ciencies del Cosmos, Universitat de Barcelona, Diagonal 647, E-08028 Barcelona, Spain}
\affiliation{$^4$ Millennium Institute for Subatomic Physics at High-Energy Frontier (SAPHIR), Fernandez Concha 700, Santiago, Chile}
\affiliation{$^5$ International Institute of Physics, Universidade Federal do Rio Grande do Norte, Campus Universit\'ario, Lagoa Nova, Natal-RN 59078-970, Brazil}
\affiliation{$^6$ Departamento de F\'isica, Universidade Federal do Rio Grande do Norte, 59078-970, Natal, RN, Brasil}

\begin{abstract}

We study the lepton-induced resonant production of color-adjoint leptons (leptogluons)  at the LHC employing the lepton parton density function of the proton. We demonstrate that this production mechanism  can be useful to extend the LHC ability to search for leptogluons beyond purely quark/gluon initiated production processes up to $\sim 3.5$ TeV leptogluon masses and ${\cal O}(1)$ TeV compositeness scales. Discerning leptogluons from scalar and vector leptoquarks is also possible in this channel, given a data sample containing the order of  $ 100$ signal events. We argue that the resonant channel can be combined with leptogluon pair and associated leptogluon-lepton productions to boost exclusion limits and discovery prospects at the LHC.
 

\end{abstract}


\maketitle

\section{Introduction}

Composite models for quarks and leptons~\cite{Pati:1974yy, Terazawa:1976xx, Neeman:1979wp, Lyons:1982jb, Harari:1982xy, Shupe:1979fv, Fritzsch:1981zh} not only contain excited states of the known particles, but also bound states carrying rather unusual quantum numbers. Among these, there are leptogluons that are color-adjoint fermions carrying non-vanishing lepton number. The existence of such states is possible if gluons and leptons contain constituents feeling the same confining force~\cite{Harari:1980ez,Squires:1980cm,Fritzsch:1981zh}.  \smallskip

Since leptogluons are charged under $SU(3)_c$, they can be pair produced at the Large Hadron Collider (LHC) through the processes $ p p \to  q \bar{q} / g g  \to \ell_8 \bar{\ell}_8 $~\cite{King:1984qk, Barger:1988pm, Baur:1985ud, Rizzo:1985dn,Rizzo:1985ud, Chivukula:1990di,Hewett:1997ce,  Mandal:2012rx, Mandal:2016csb}, where we denote charged leptogluons by $\ell_8$ with $\ell=e$, $\mu$ or $\tau$. In addition to this channel, they can also be singly produced in association with a charged lepton $ p p \to q\bar{q} / g g\to \llg \ell$~\cite{Mandal:2012rx, Mandal:2016csb}.  Moreover, leptogluons also contribute to the Drell-Yan process via their higher-dimension interaction with gluon-lepton pairs~\cite{Jelinski:2015epa}. \smallskip

In this work, we study the production of leptogluons as an  $s$-channel resonance $ p p \to g e \to e_8 \to g e$, where the electrons in the proton are described by  a parton distribution function~\cite{Bertone:2015lqa, Buonocore:2020nai}. This process is analogous to resonant leptoquark production via quark-electron collisions in hadron colliders~\cite{Ohnemus:1994xf, CiezaMontalvo:1998sk, PhysRevLett.125.231804}. We demonstrate that this channel extends the LHC reach to search for leptogluons depending on the strength of their non-renormalizable interactions. Moreover, we also study how to distinguish between the production of leptogluons and scalar or vector leptoquarks. \smallskip

Presently there are few experimental limits on leptogluons. These exotic states have been searched at the HERA $ep$~collider through their resonant production in the $s$-channel~\cite{H1:1993vsn}. Stable charged leptogluons were also searched by the CDF collaboration at the Tevatron~\cite{CDF:1989jud}, leading to $m_{e_8} > 86 $ GeV at 95\% CL. The JADE collaboration at PETRA studied final states containing jets and leptons and  excluded leptogluons with masses in the 100-200 GeV range for compositeness scales $\Lambda $ in the range 1--2 TeV. Presently, the most stringent limits originate from phenomenological analyses based on the leptogluon  decays into gluon--lepton pairs through non-renormalizable operators~\cite{Goncalves-Netto:2013nla,Mandal:2016csb}. Since the production mechanisms and decays of leptoquarks and leptogluons are similar, it is possible to translate  experimental limits on the latter to leptogluons, leading to $m_{\ell_8} > 1.2$ TeV at 95\% CL~\cite{Goncalves-Netto:2013nla}. This limit was  obtained from a recast of double scalar leptoquark searches at the 7 TeV LHC by the CMS Collaboration in the $e^+e^- jj$ channel~\cite{CMS:2012iln}. More recently, a recast of the 8 TeV LHC data was performed, extending that limit from double production to $\sim 1.55$ TeV~\cite{Mandal:2016csb}. \smallskip

Our work is organized as follows. In section~\ref{sec:analysis}, we give the details of our phenomenological analysis,  and  we present the exclusion limits and the discovery prospects of leptogluons in this channel in section~\ref{sec:results}. Section~\ref{sec:discern} is devoted to discerning leptogluons from scalar and vector leptoquarks, and  we present our conclusions in section~\ref{sec:conclude}. \smallskip

\section{Analysis framework}
\label{sec:analysis}

The Lagrangian density describing the interaction among leptogluons, gluons, and charged leptons is given by
\begin{equation}
     \mathcal{L}_{int} =  - g_s f^{abc} \lgx^a \gamma^\mu \llg^b A_\mu^{c} + \frac{g_s}{2\Lambda} \lgx^a \sigma^{\mu\nu} G^a_{\mu\nu}(a_LP_L+a_RP_R) \ell + \hbox{h.c.}
\label{eq:int}
\end{equation}
where $\ell$ and $A_{\mu}^c$ denote the fields of  a standard model) charged lepton and the gluon, respectively\footnote{Here, we do not consider neutral leptogluons which couple to neutrino-gluon pairs}. The gluon field strength tensor is defined as $G_{\mu\nu}^a=\partial_\mu A^a_\mu+\partial_{\nu} A_{\mu}^a+g_s f^{abc}A^b_{\mu} A^c_{\nu}$, $g_s$ is the strong coupling and $f^{abc}$ is the $SU(3)_C$ structure constant. The first term in Eq.~(\ref{eq:int}) is just the $SU(3)_C$ gauge interaction of a colored fermion. On the other hand, the second term in the above equation is the lowest dimension non-renormalizable operator generated by the confining strong interaction that is parametrized by the compositness scale $\Lambda$ and leptogluon coupling to left-handed (right-) leptons $a_L$ ($a_R$).  We also assume that leptogluons conserve lepton number, which implies the existence of three different charged leptogluon fields, {\em i.e.}  $e_8$, $\mu_8$ and $\tau_8$. \smallskip

\begin{figure}[!hb]
    \centering
     \includegraphics[scale=0.5]{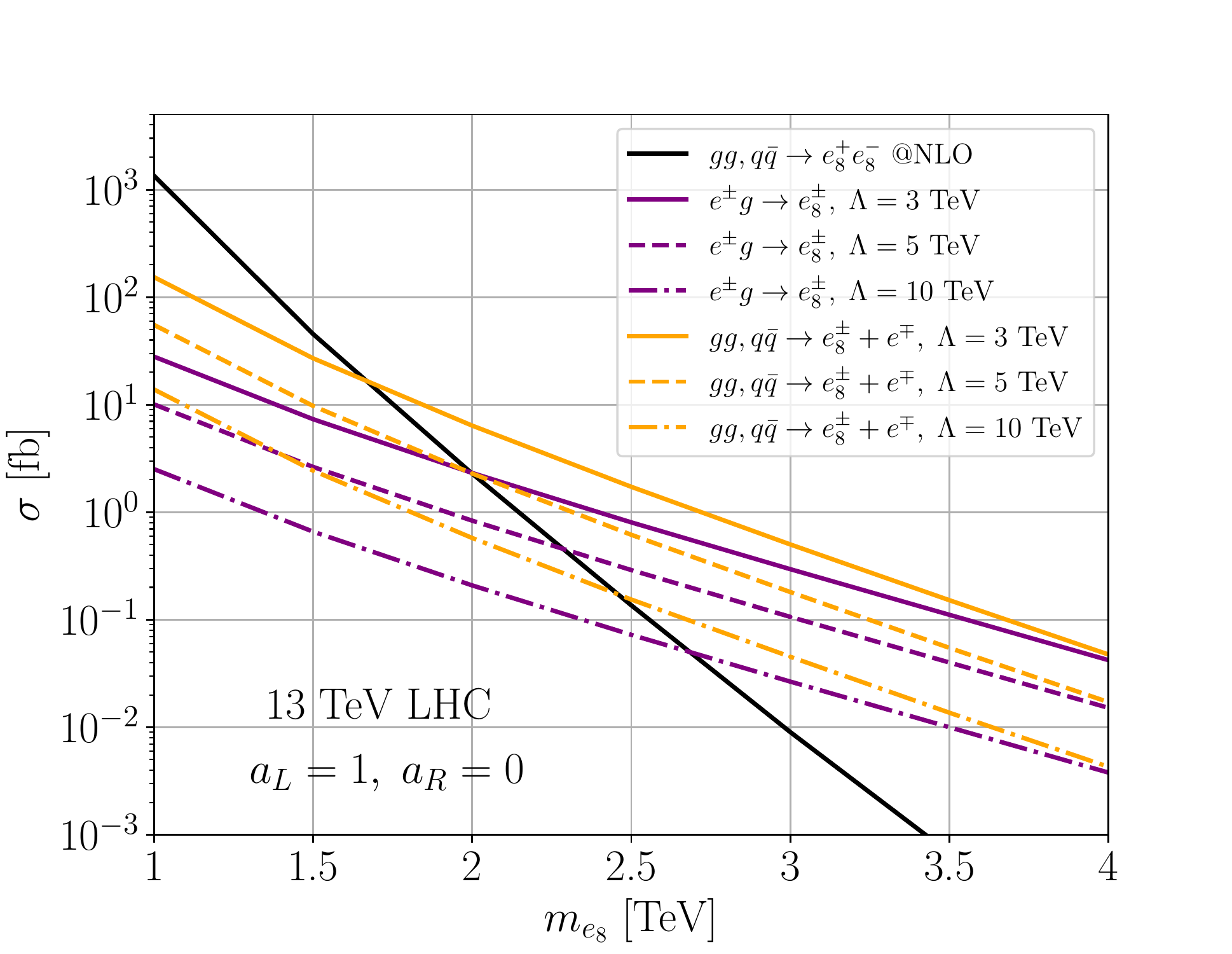}
    \caption{ Leptogluon production cross sections at the 13 TeV LHC. We denote by the solid black line the double leptogluon $e_8\bar{e_8}$ production cross section and by the orange lines  the associated  production $e_8 e$, while the purple lines stand for $s$-channel resonant production. The leptogluon pair  production was evaluated at QCD NLO, and the other mechanisms at LO for the couplings indicated in the figure. }
\label{xsec}
\end{figure}

 The interactions in Eq.~\eqref{eq:int} allow for the double leptogluon production ($gg(q\bar{q})\to \llg\lgx$) via gauge interactions, as well as for the production of a leptogluon in association with a lepton ($gg(q\bar{q})\to \llg\ell$) through its non-renormalizable interaction to gluon-lepton pairs. Moreover,  since protons also contain leptons, it is possible to have  single leptogluon resonant production $g\ell(\bar{\ell})\to \llg(\lgx)$~\cite{Buonocore:2020nai,Buonocore:2020erb}. In Figure~\ref{xsec} we depict the cross sections for these mechanisms as a function of the leptogluon $e_8$ mass. We evaluated the double leptogluon production at next-to-leading order (NLO)~\cite{Goncalves-Netto:2013nla}, while the single and resonant productions were evaluated at leading order. As we can see from this figure, the resonant and associated leptogluon productions dominate the cross section at  large leptogluon masses, with the resonant mechanism leading to the largest contribution for heavy leptogluons.  In fact, the single production rate surpasses the double one if $\Lambda=3~(10)$ TeV for leptogluon masses larger than $m_{e_8}\sim 1.5~(3)$ TeV with $a_L=1,\; a_R=0$. This is a conservative result since higher-order QCD corrections are expected to increase the resonant and associated leptogluon productions. On the other hand, our LO estimate might be justified once we will veto any hadronic activity beyond a leading jet as we discuss ahead. \smallskip

Considering the interactions given in Eq.~(\ref{eq:int}), leptogluons decay into a lepton and a gluon with a decay width
\begin{equation}
    \Gamma_{\llg \to \ell g} = \frac{\alpha_s}{4} \frac{m_{\llg}^3}{\Lambda^2} (a_L^2 + a_R^2) \; .
    \label{eq:width}
\end{equation}

The signal of the resonant leptogluon production is the presence of a charged lepton and a jet with the lepton-jet invariant mass peaking around the leptogluon mass. The main background sources in the case  are  QCD multijets, $W+j$, and $Z+j$, similar to the resonant production of leptoquarks~\cite{PhysRevLett.125.231804}. Subdominant background sources are  diboson ($WW$ and $WZ$) and single top productions.\smallskip

For a large momentum fraction of the proton carried by the lepton, the evolution equations for the lepton parton distribution function (PDF)  can be calculated perturbatively, analogously to the photon PDF, and  it can be fitted from data. Here, we use the lepton PDF  derived in Ref.~\cite{Buonocore:2020nai}, and that can be found in the LUXlep PDF set of  \texttt{LHAPDF}~\cite{Buckley:2014ana}. \smallskip

 We simulate the signal process $g+ e^-(e^+)\to e_8^- ~(e_8^+)\to g+ e^- ~(e^+)$ at leading order with a modified version of \texttt{MadGraph5}~\cite{Alwall:2014hca} to handle initial state leptons, fixing $\mu_R=\mu_F=m_{e_8}$ as the renormalization and factorization scales, respectively. We do not take muons into account to compare our results directly to those of Refs.~\cite{Goncalves-Netto:2013nla,Mandal:2016csb} but assuming that leptogluons have universal couplings to all leptons would increase the production cross section and double the branching ratio extending the reach of the LHC for this channel.
 The partonic events were showered and hadronized with \texttt{Pythia8}~\cite{Sjostrand:2007gs}, and detector effects were simulated with \texttt{Delphes3}~\cite{deFavereau:2013fsa}. \smallskip

 We used the anti-$k_t$ algorithm with radius $\Delta R=0.4$ to reconstruct jets with \texttt{Fastjet}~\cite{Cacciari:2011ma}. Electrons and muons were considered isolated if no net activity with transverse momentum in excess of 10 GeV was found around a cone of $\Delta R=0.5$ around the lepton momenta. \smallskip
 
\begin{figure}[!ht]
    \centering
     \includegraphics[scale=0.5]{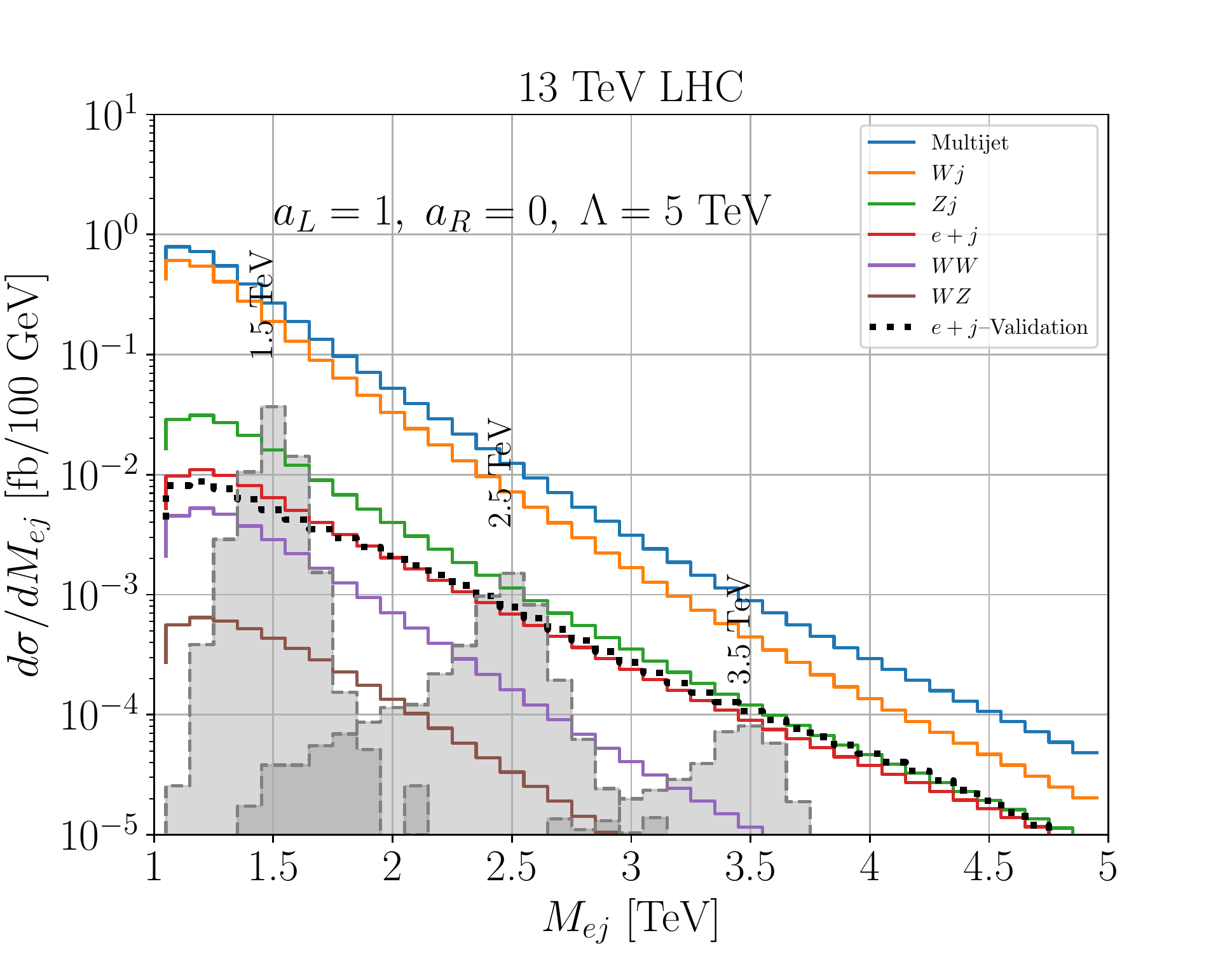}
    \caption{Electron-jet  invariant mass distribution for the main backgrounds to the resonant leptogluon production. We also present the expected signal distributions for three $m_{e_8}$ as indicated. }
\label{mlj}
\end{figure}

 Our search analysis follows the resonant leptoquark one presented in Ref.~\cite{Buonocore:2020erb} closely. Just like the leptoquark case, we required the following basic cuts on the leptogluon signal 
\begin{eqnarray}
 p_{T_e} & > & 500\;\hbox{GeV},\;\; |\eta_e|<2.5, \\
 p_{T_j} & > & 500\;\hbox{GeV},\;\; |\eta_j|<2.5, \\
 & & \not \!\! E_T < 50\;\hbox{GeV}\; .
 \label{eq:cuts}
 \end{eqnarray}
While the transverse momentum cut favors a heavy resonance in excess of 1 TeV mass, the missing transverse energy helps to keep events with no final state neutrinos or missing leptons and/or misreconstructed jets. This very same cut strategy is efficient in our case because the scalar leptoquarks and leptogluon lower limits based on pair production searches are similar, around $1.48$ TeV~\cite{ATLAS:2020xov, CMS:2018ncu, Schmaltz:2018nls} and $1.55$ TeV~\cite{Goncalves-Netto:2013nla, Mandal:2016csb}, respectively. \smallskip

In order to further suppress the multijet and top quark backgrounds, we veto additional jets with $p_T>30$ GeV within $|\eta_j|<2.5$ and additional leptons with $p_T>7$ GeV and $|\eta_\ell|<2.5$~\cite{Buonocore:2020erb}.  A final requirement to isolate the resonant signal is $|M_{e j}-m|<\delta_m$, where $M_{e j}$ is the lepton-jet invariant mass. The parameters $m$ and $\delta m$ are adjusted  to maximize the Azimov statistical significance~\cite{Cowan:2010js}. \smallskip

Instead of simulating the backgrounds again, we took the results from Ref.~\cite{Buonocore:2020erb}, which depicts the $M_{\ell j}$ distribution for the backgrounds in 100 GeV bins as shown in Figure~\ref{mlj} after applying the above kinematic requirements. In order to reproduce the background distributions of Ref.~\cite{Buonocore:2020erb} as accurately as possible, we performed a fit of the histograms with the functional form $e^{-a} M_{\ell j}^{P(M_{\ell j})}$, where $P(M_{\ell j})=\sum_{n=1}^6 a_n (\ln\ M_{\ell j})^n$. Moreover, in order to validate our simulations, we generate particle-level events for the SM background component $e+q\to e+q$ initiated by an electron-quark collision exchanging a $Z$-boson/photon in the $t$-channel. Our results, depicted as the dotted-black distribution in Figure~\eqref{mlj}, show a reasonable agreement with the $e+j$ background simulation of Ref.~\cite{Buonocore:2020erb}. We also depict the expected lepton-jet invariant mass spectrum for three representative values of the leptogluon mass.\smallskip

Compared to  scalar leptoquarks, leptogluons tend to be   wider resonances once their total width scales as $m_{e_8}^3/\Lambda^2$, while the scalar leptoquark one scales linearly with its mass. Moreover, we expect that the jet multiplicity in  resonant leptogluon production is higher than in the leptoquark case. These two features favor a higher selection efficiency for leptoquarks than leptogluons in the analysis of Ref.~\cite{Buonocore:2020erb}; therefore, the leptogluon prospects for detection are expected to be reduced with respect to the leptoquark ones. Yet, as we will see in the next section, the resonant leptogluon search is able to extend the limits of the current pair production searches.\smallskip

\section{LHC Exclusion and discovery prospects}
\label{sec:results}

In our analyses, we used the Azimov statistical significance ($Z_A$) to obtain exclusion limits and discovery regions
\begin{equation}
 Z_A(s,b,\sigma_b) = \left[2\left((s+b)\ln\left[\frac{(s+b)(b+\sigma_b^2)}{b^2+(s+b)\sigma_b^2}\right]-\frac{b^2}{\sigma_b^2}\ln\left[1+\frac{s\sigma_b^2}{b(b+\sigma_b^2)}\right]\right)\right]^{1/2} \;,
\end{equation}
where $s$ and $b$ are the number of signal and background events after all cuts, $ \sigma_b^2 = \varepsilon_b b$ and with $\varepsilon_b$ being the systematic uncertainty of the total background rate. The optimization of the lepton-jet invariant mass cut  for maximum signal significance is performed by adjusting the $m$ and $\delta_m$ parameters defined above such that 
\begin{equation}
 N_\sigma = \underset{m,\delta_m}{\mathrm{argmax}}\; Z_A(s(m,\delta_m),b(m,\delta_m),\varepsilon_b) \;.
\end{equation}

\begin{figure}[!ht]
    \centering
     \includegraphics[scale=0.5]{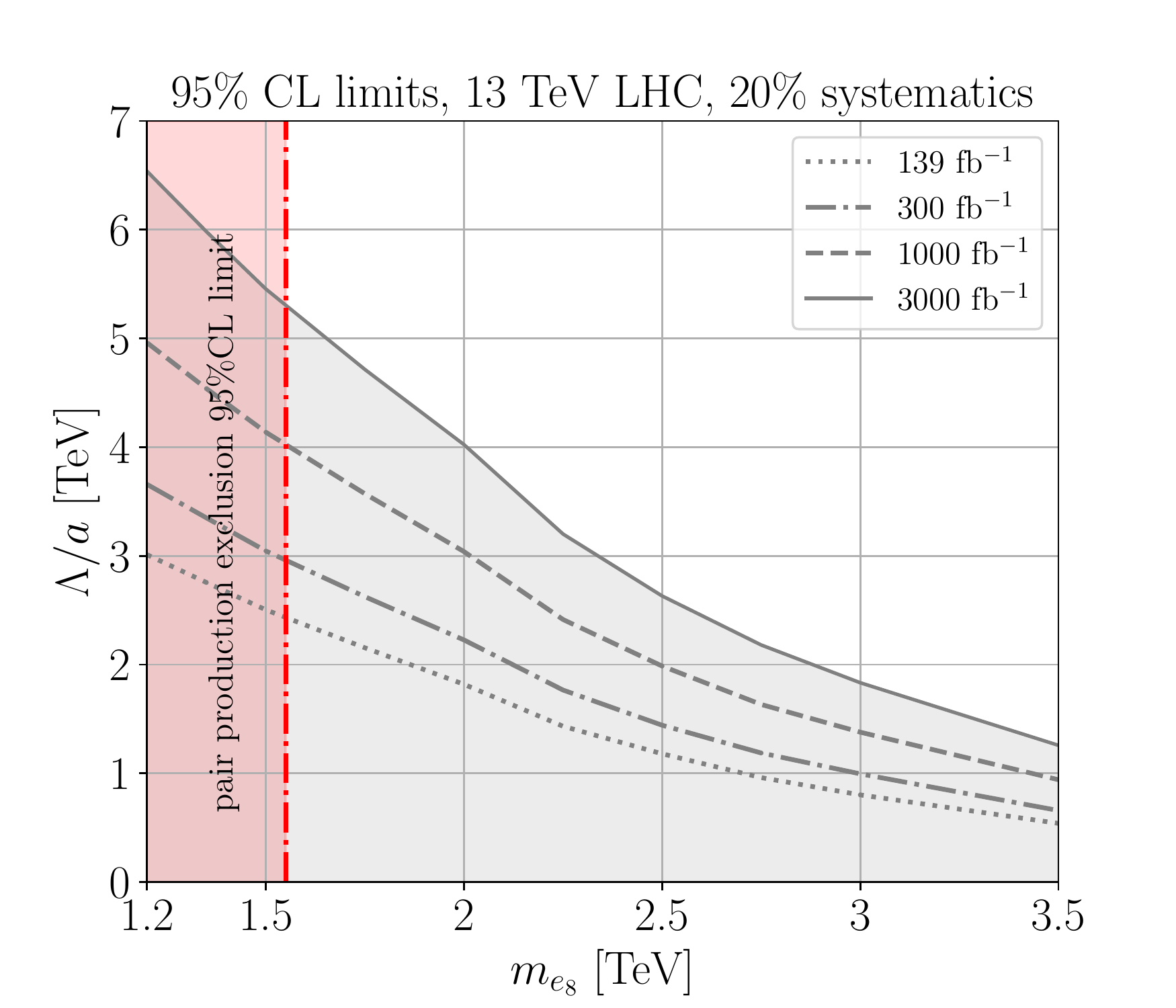}
        \caption{95\% CL exclusion limits in the $m_{e_8} \times \Lambda/a$ plane due to the resonant leptogluon search considering systematic uncertainties of  20\%. The excluded regions are indicated by gray shaded regions for several integrated luminosities as indicated. }   
\label{results:excl}
\end{figure}

Figure~\ref{results:excl} depicts the  95\% CL exclusion region in the $m_{e_8} \times a_L/\Lambda$ parameter space due to the resonant leptogluon search assuming a conservative systematic uncertainty of   $\varepsilon_b=20$\%. For definiteness, we considered only leptogluons coupling to electrons and set $a_R=0$. The vertical dash-dotted line indicates the pair production 95\% CL limit, which exclude $e_8$ leptogluons lighter than $\sim 1.55$ TeV irrespective of the couplings $a_{L/R}/\Lambda$~\cite{Goncalves-Netto:2013nla, Mandal:2016csb}. The shaded regions below the gray solid, dashed, dot-dashed, and dotted curves represent the portions of the parameter space excluded at that confidence level if the number of observed events is the one predicted by the SM for 139, 300, 1000, and 3000 fb$^{-1}$ respectively. 
%
%
For large leptogluon masses, its production cross section is dominated by the resonant and associated mechanisms; therefore, we can obtain stronger limits by combining these channels. 

\begin{figure}[!ht]
    \centering
     \includegraphics[scale=0.5]{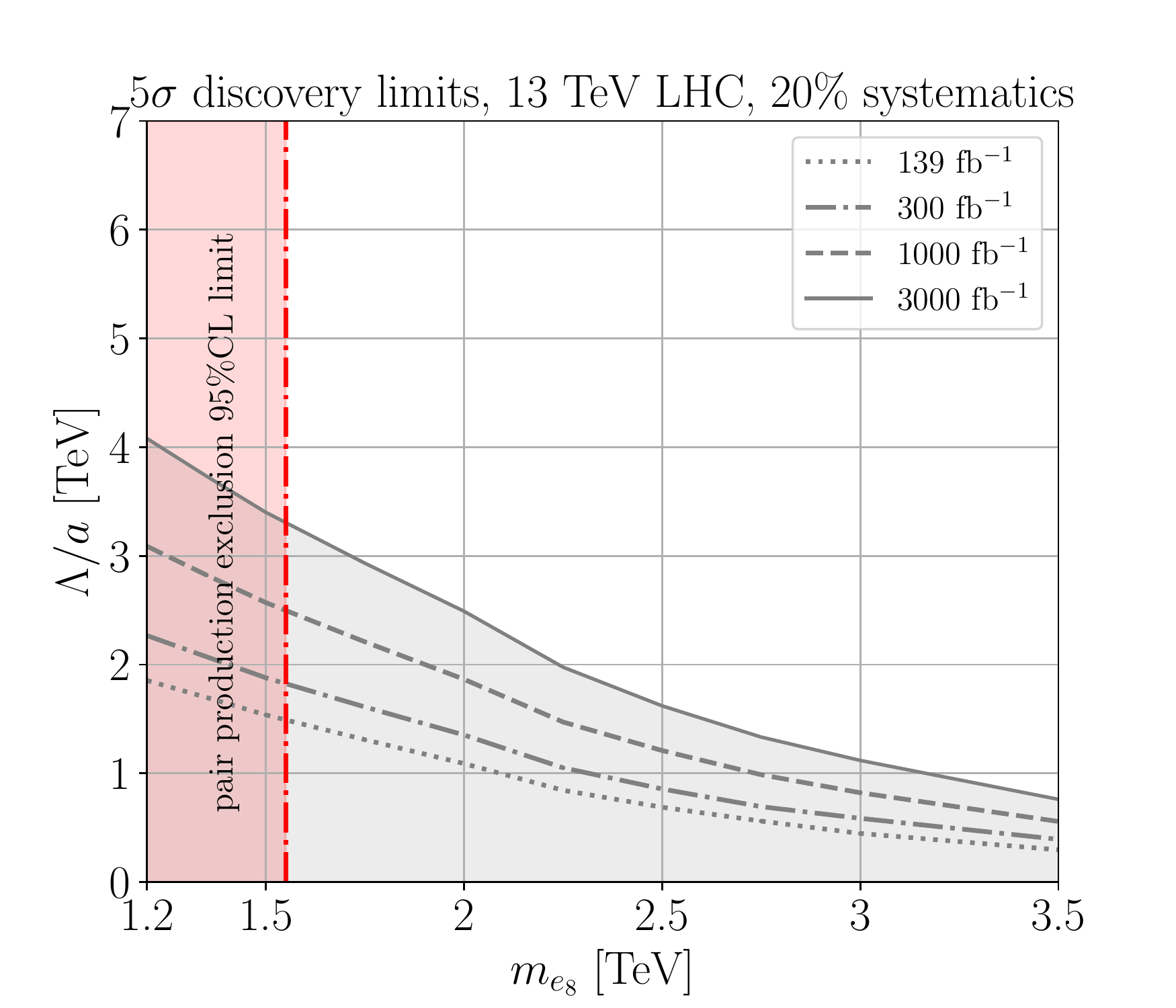}
    \caption{The $5\sigma$ discovery of leptogluons in the resonant channel  is possible in the shaded area below the lines for several luminosities. Here, we assumed a $20$\% systematic uncertainty, and the  integrated luminosities are as indicated in the figure.  }
\label{results:disc}
\end{figure}

In Figure \ref{results:disc}, the gray regions below the solid, dashed, dot-dashed and dotted lines  mark  the parameter-space region where a $5\sigma$ leptogluon discovery is possible in the resonant channel for the indicated  integrated luminosities. Once again, we considered 20\% systematic uncertainties in our analysis. In the high leptogluon mass, its production cross section is dominated by the associated and resonant mechanisms, see Figure~\ref{xsec}; therefore, the combination of these channels is desirable to increase the LHC discovery reach. Moreover, we have verified that the results are rather insensitive to the assumed systematic uncertainties since statistical errors are dominant. \smallskip

It is also important to notice that our analysis can be optimized for heavier leptogluons by adjusting the kinematic cuts, as performed in Ref.~\cite{Mandal:2016csb}. In particular, the jet veto performed to suppress the backgrounds in Ref.~\cite{Buonocore:2020erb}, and adopted in our analysis, is more penalizing in the case of leptogluons compared to leptoquarks. An optimization of cuts or the classification of signal and background events with machine learning algorithms might help to increase the signal significance. \smallskip
 
\section{Discriminating  Leptogluons from  Leptoquarks}
\label{sec:discern}

Leptoquarks and leptogluons lead to the same signal topologies at the LHC. Therefore,  if a resonance in the lepton+jet channel is discovered above the SM backgrounds, it is mandatory to study kinematical distributions to discriminate between these possible candidates due to their different spins. Since their single productions depend on an unknown coupling, it is not possible to tell them apart using just the observed production cross section.  Therefore, we tested the leptogluon hypothesis against a scalar or vector leptoquark one relying just on the shape of their kinematic distributions but assuming the same number of events in a conservative analysis. \smallskip

In order to study how to differentiate between these possible states, we considered one scalar and one vector leptoquark states whose interaction Lagrangians are given by~\cite{Buchmuller:1986zs}
\begin{equation}
    \mathcal{L}_{\rm LQ} =  h_{2L} R_2^T \bar{u}_R i\tau_2 L_L + h_{1L} U_{1\mu}  \bar{Q}_L \gamma^\mu L_L 
\end{equation}
where  $Q_L$ and $L_L$ stand for quark and lepton doublets and $u_R$ for the up quark singlets. Here, the coupling of the scalar (vector) leptoquark $R_2$ ($U_{1\mu}$) is $h_{2(1)L}$ and   $\tau_j$ are the Pauli matrices. \smallskip

\begin{figure}[!ht]
    \centering
     \includegraphics[scale=0.285]{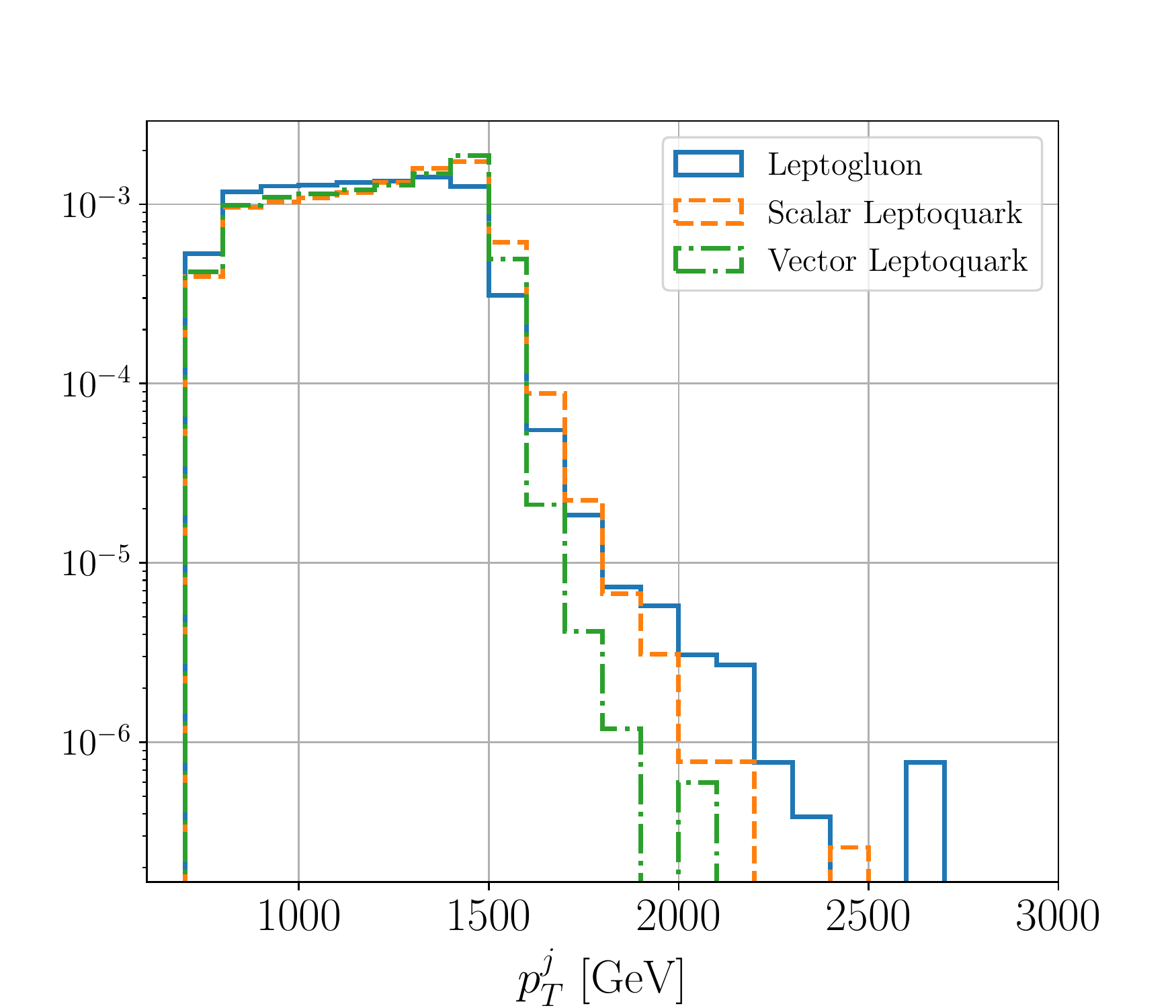}
     \includegraphics[scale=0.285]{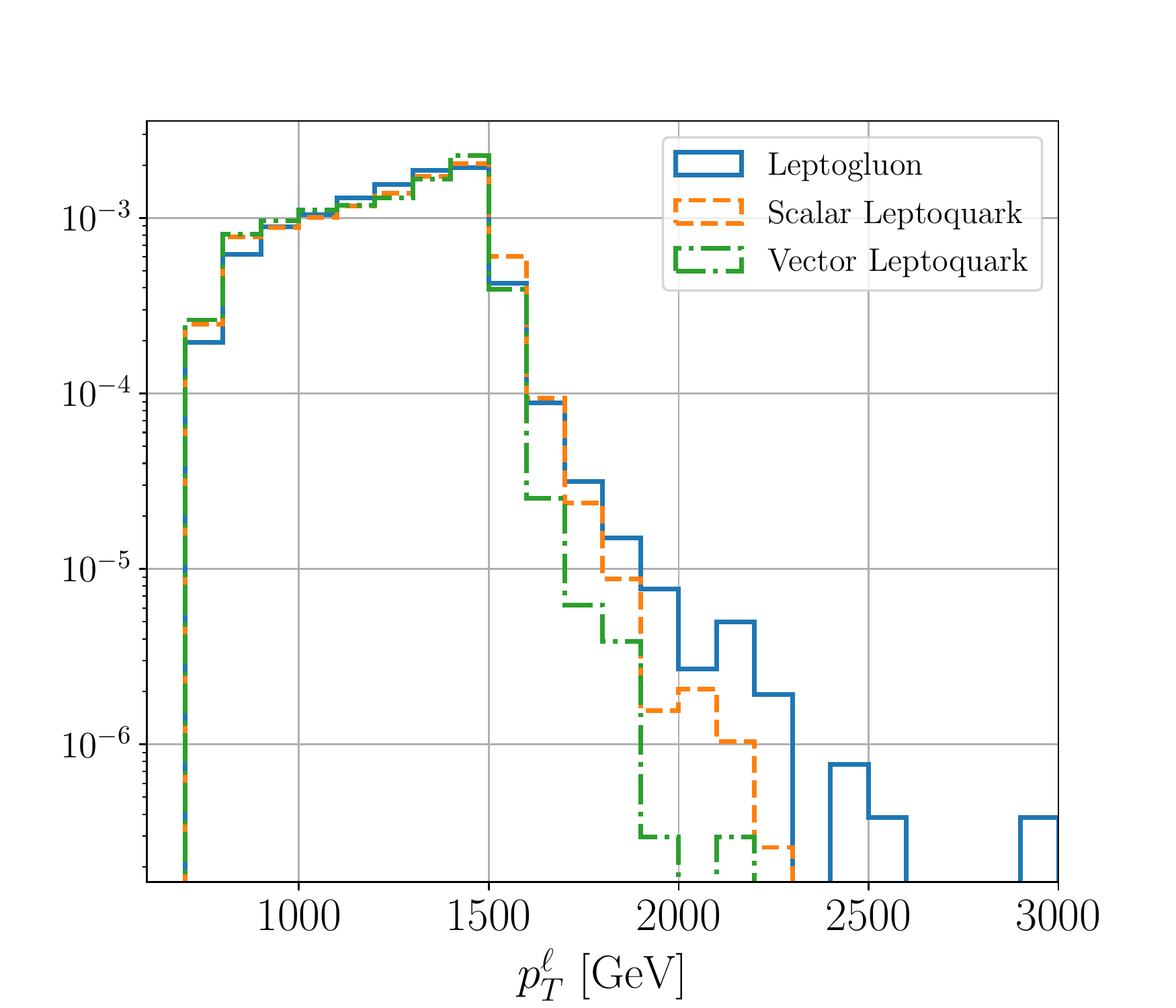}
     \includegraphics[scale=0.285]{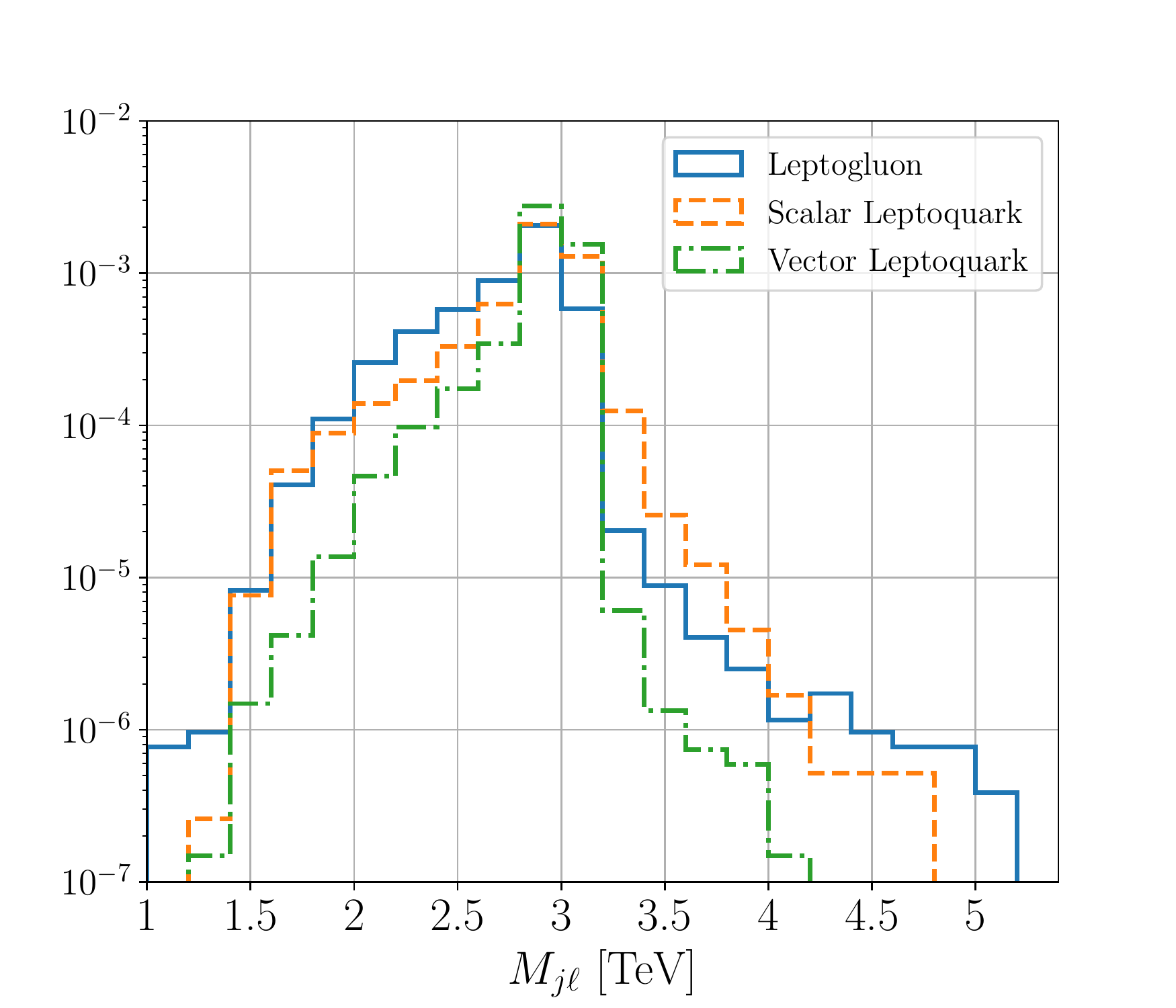}\\
     \includegraphics[scale=0.285]{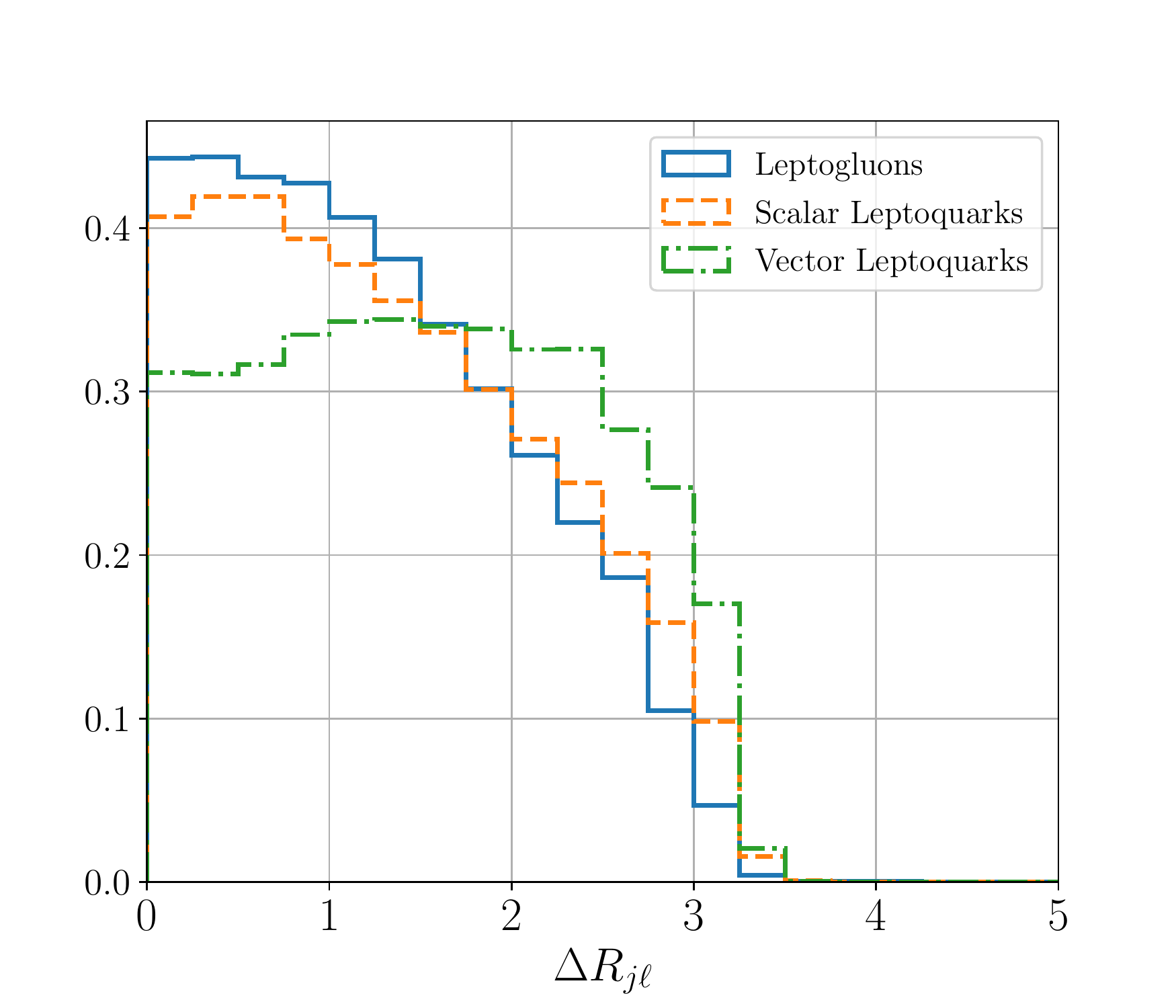}
     \includegraphics[scale=0.285]{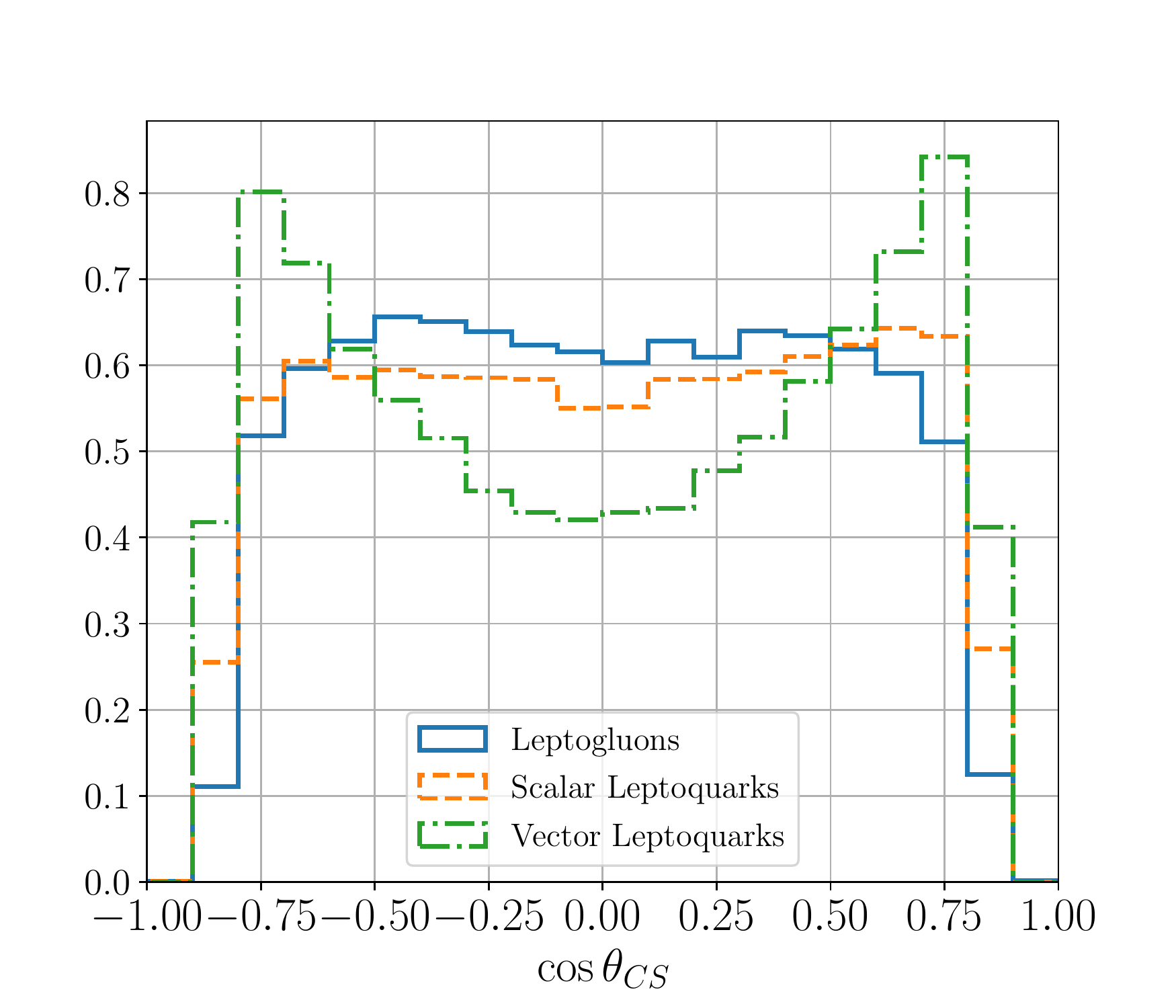}
     \includegraphics[scale=0.285]{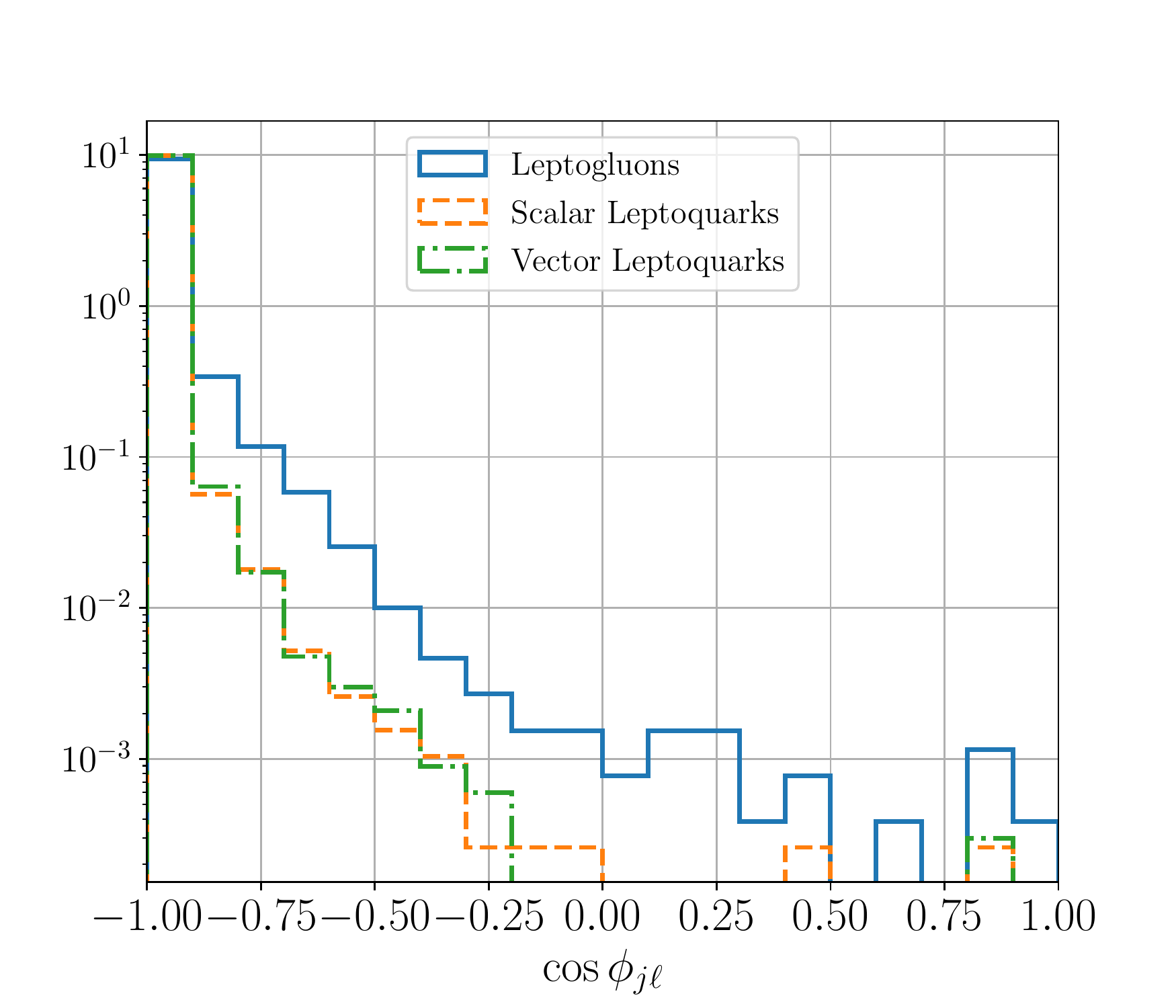}
    \caption{Normalized kinematic distributions for scalar and vector leptoquarks, as well as leptogluons for the resonant production of a 3 TeV state. The upper left, center and right panels depict the transverse momentum of the leading lepton ($p_{T,\ell}$),  the transverse momentum of the leading jet ($p_{T,j}$), and the lepton-jet invariant mass ($M_{j\ell}$) distributions, respectively. The lower left, center, and right panels display the lepton-jet separation ($\Delta R_{j\ell}$),  the cosine of the angle between the lepton and the jet in Collins-Soper frame ($\cos\theta_{j\ell}$) and the azimuthal angle between jet and lepton ($\cos\phi_{j\ell}$) spectra.}
\label{fig:dists}
\end{figure}

In order to differentiate the signal of leptogluons and leptoquarks, we studied the following normalized kinematical distributions, illustrated in Figure~\ref{fig:dists} for a 3 TeV new state:  the transverse momentum of the leading lepton $p_{T,\ell}$  (upper left panel) the transverse momentum of the leading jet $p_{T,j}$ (upper center panel), the lepton+jet invariant mass $M_{j\ell}$ (upper right panel), the distance between the lepton and the jet in the $\phi$-$\eta$ plane $\Delta R=\sqrt{(\eta_\ell-\eta_j)^2+(\phi_\ell-\phi_j)^2}$ (lower left panel), the cosine of the angle between the lepton and the jet in Collins-Soper frame~\cite{ATLAS:2015zhl,PhysRevD.16.2219} (lower center panel) 
\begin{equation}
     |\cos\theta_{CS}| = \frac{|\sinh(\Delta\eta_{j\ell})|}{\sqrt{1+(p_T^{j\ell}/M^2_{j\ell})^2}}\frac{2p_T^\ell p_T^j}{M_{j\ell}^2},
\end{equation}
and  the cosine of the azimuthal angle between the leading lepton and jet (lower right panel). As we can see from the panels in this figure, it is difficult to distinguish clearly between the three possible new states using just one distribution. \smallskip

%

\begin{figure}[!ht]
    \centering
     \includegraphics[scale=0.285]{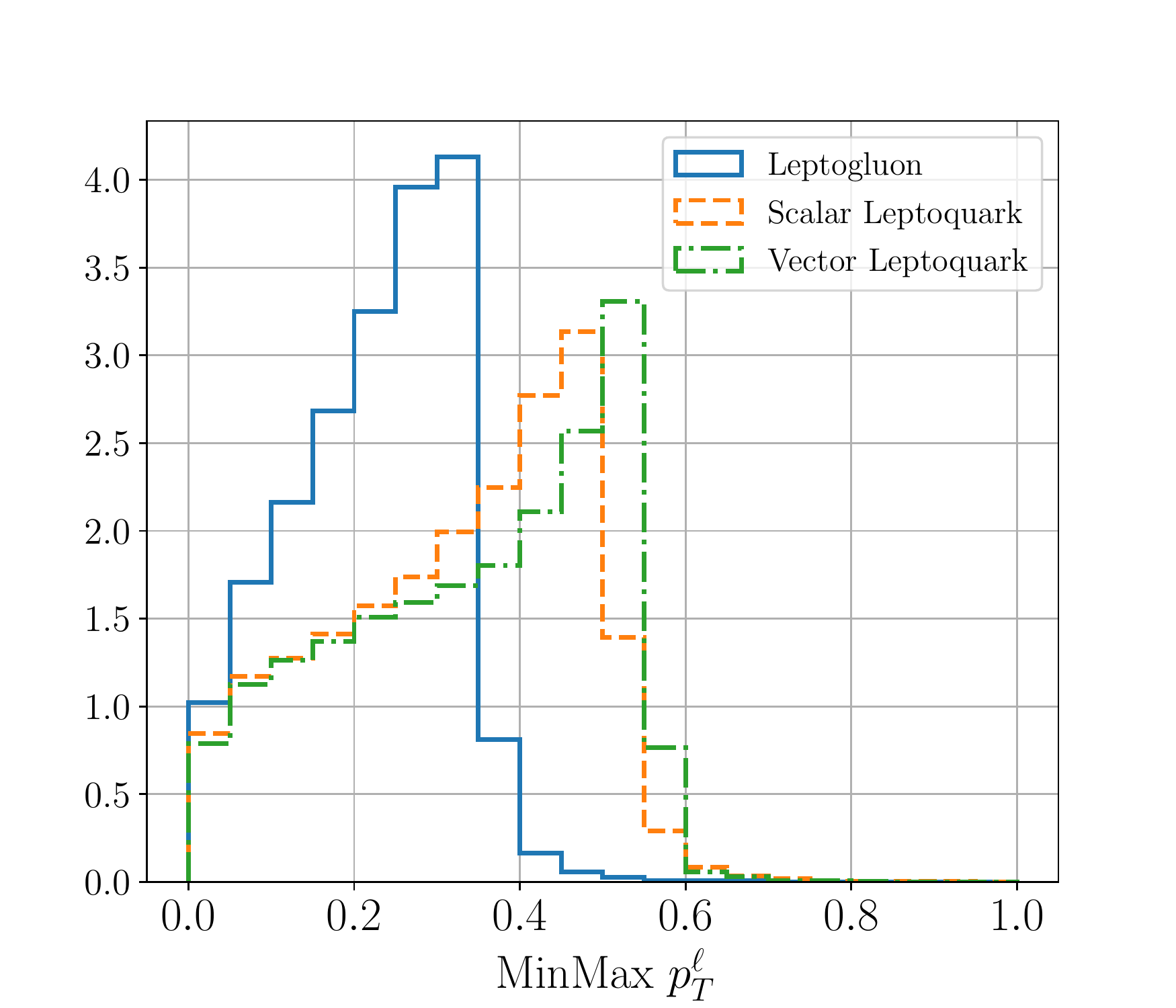}
     \includegraphics[scale=0.285]{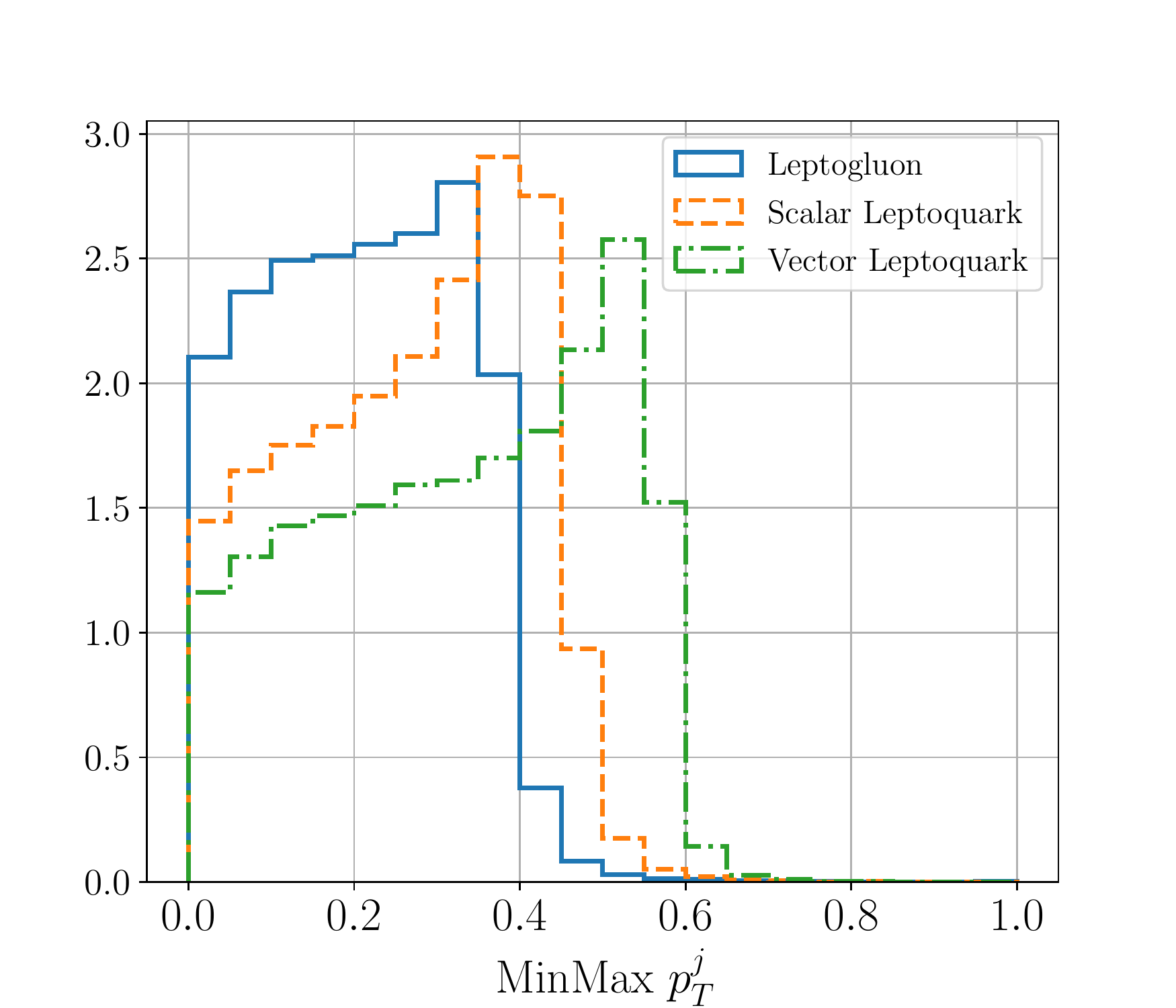}
     \includegraphics[scale=0.285]{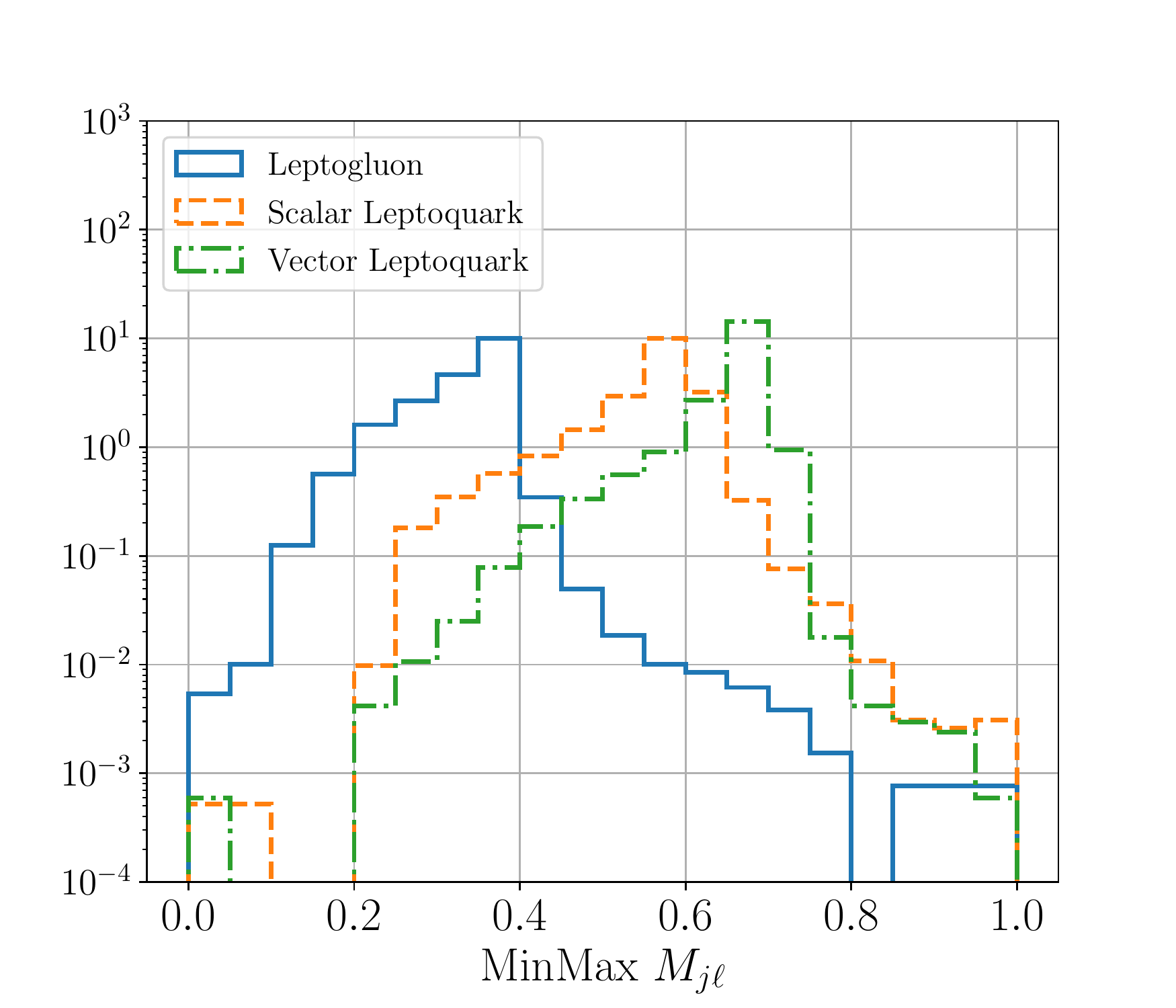}\\
     \includegraphics[scale=0.285]{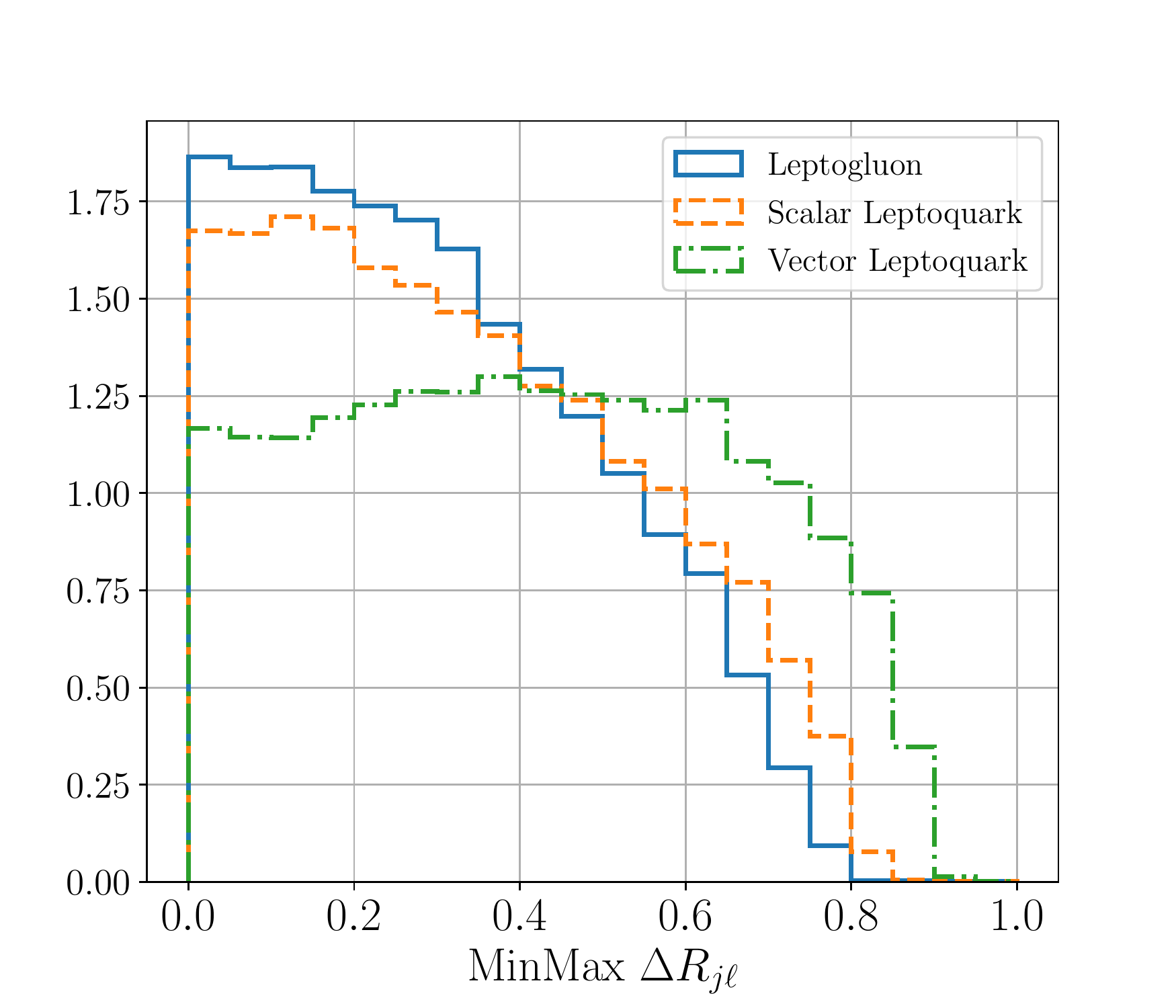}
     \includegraphics[scale=0.285]{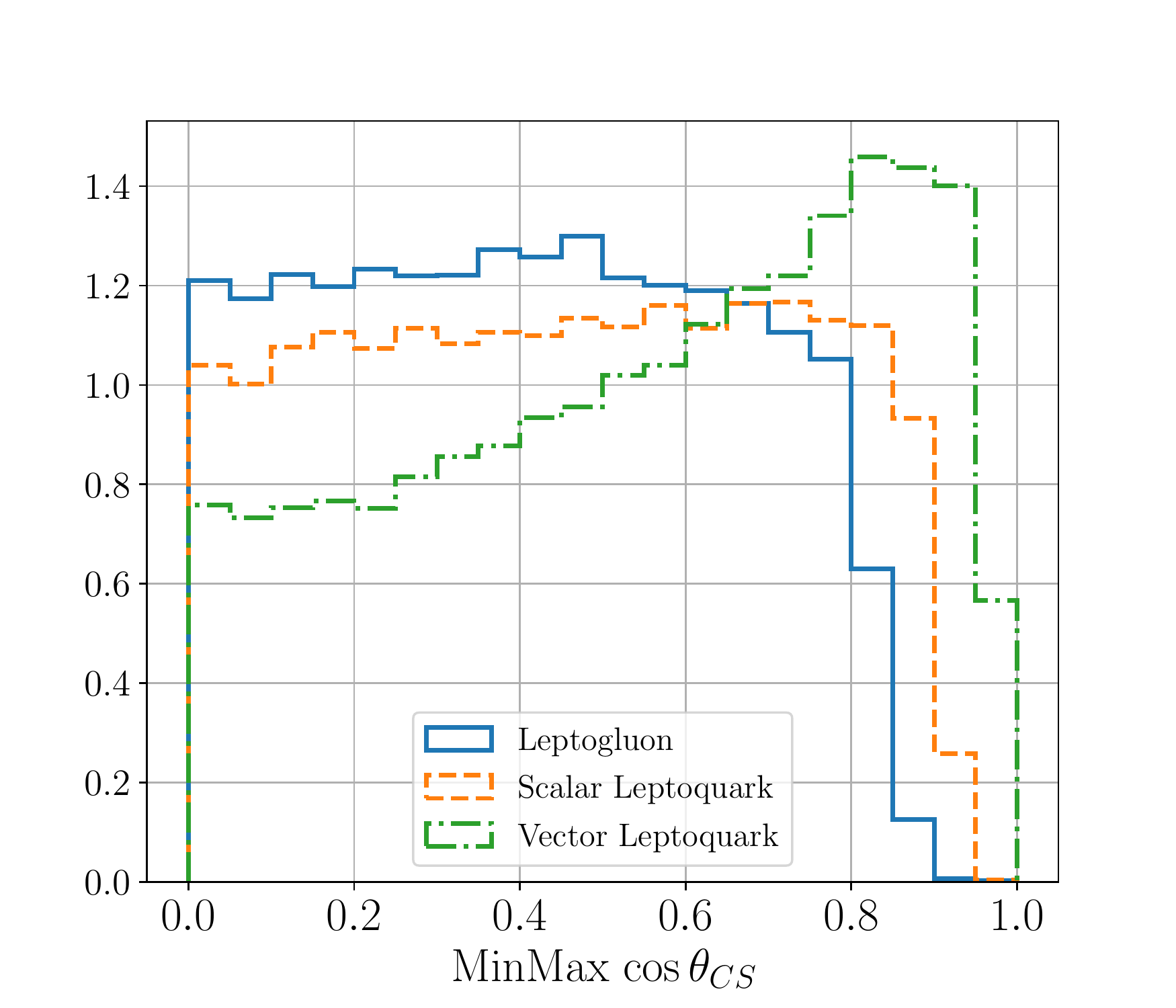}
     \includegraphics[scale=0.285]{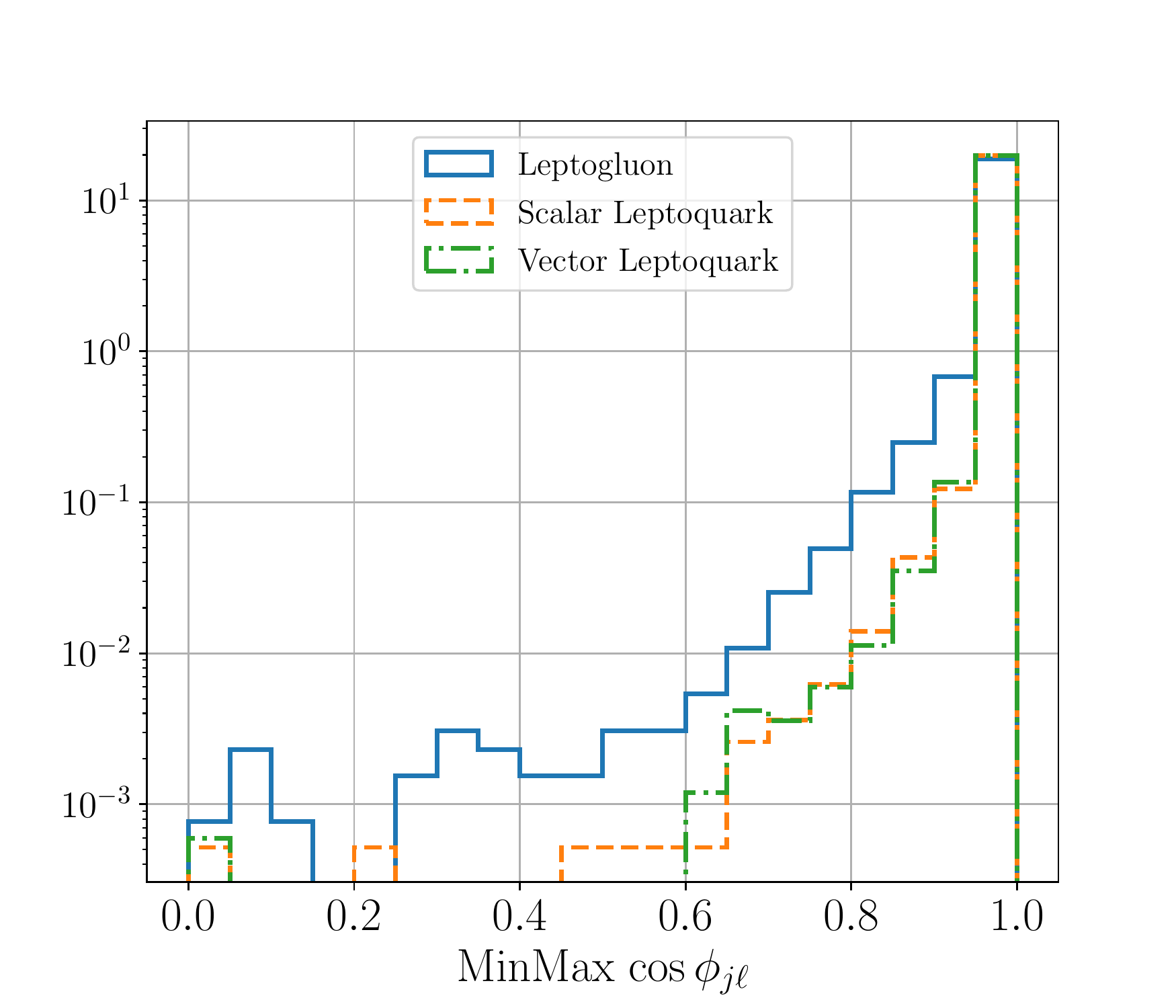}
    \caption{{\bf MinMax} transformed distributions of the kinematic distributions in Figure~\ref{fig:dists}.}
\label{fig:dists-minmax}
\end{figure}

 The salient features of a kinematic variable $x$ can be enhanced by a {\bf MinMax} transformation  $x\to (x-\min(x))/(\max(x)-\min(x))$. This transformation also makes it easier to compute the binned log-likelihood ratio by restricting the range of the variables to $[0,1]$. The {\bf MinMax} transformed kinematic  variables are shown in Figure~\ref{fig:dists-minmax}. In particular, we see that the mass dimension variables, the transverse momenta, and the lepton-jet invariant mass have their peaks shifted compared to the original distributions of Figure~\ref{fig:dists}, making the distinction among the hypotheses clearer. The larger shift towards zero of the leptogluon distributions compared to leptoquark ones occurs because the leptogluon distributions are harder/wider due to enhanced QCD radiation (see Figure~\ref{fig:dists}), making the difference between the maximum and minimum of the distributions larger in the case of leptogluons.  
 On the other hand, the angular variables are more suitable for discriminating between scalar and vector leptoquarks. \smallskip

Using the {\bf MinMax}  distributions, we calculate the binned log-likelihood ratio for the two hypotheses $e_8 \times R_2$ and $e_8 \times U_1$ as
\begin{equation}
    \lambda = \sum_{i=1}^6\sum_{k=1}^{n_i} \left[s^{(lg)}_{ik}-s^{(lq)}_{ik}-d_{ik}\ln\left(\frac{s^{(lg)}_{ik}}{s^{(lq)}_{ik}}\right)\right],
\end{equation}
 where $n_i$, the number of bins of the $i$-th distribution, is chosen in such a way that no bins of the histogram are empty.  $s^{(h)}_{ik}$ is the number of events in the $k$-th bin of the $i$-th distribution when $h=lg,lq$ is the leptogluon and scalar(vector) leptoquark hypothesis, respectively. To estimate the log-likelihood ratio distribution of each hypothesis, we simulate 50k pseudo-experiments where $d_{ik}\sim p(x_{ik}|s^{(h)}_{ik})$ from the Poisson distribution of mean $s^{(h)}_{ik}$ represents the observed data for $x_{ik}$ observations in $k$-th bin of the $i$-th distribution when the true hypothesis is taken as $h=lg,lq$. 

\begin{figure}[!ht]
    \centering
     \includegraphics[scale=0.45]{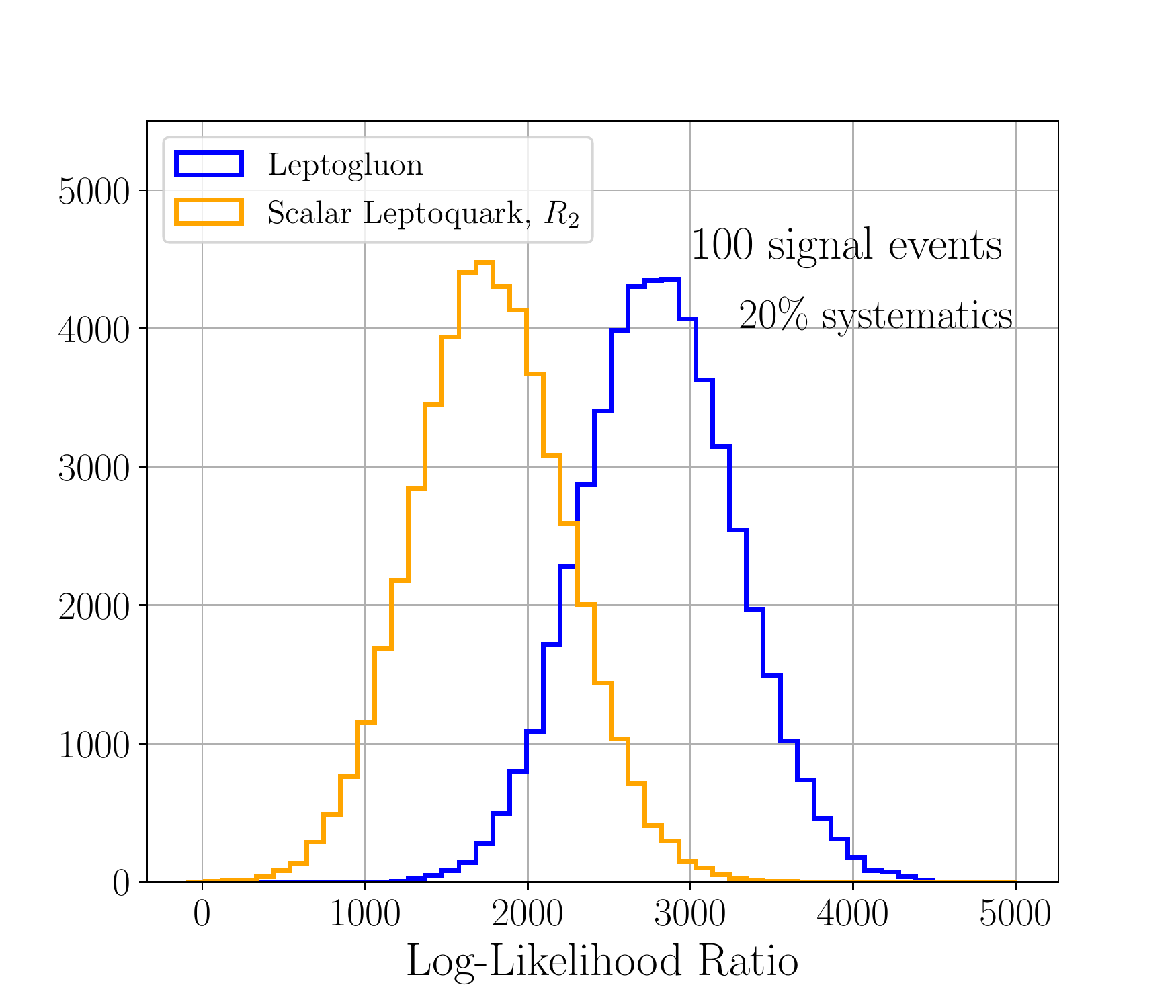}
    \includegraphics[scale=0.45]{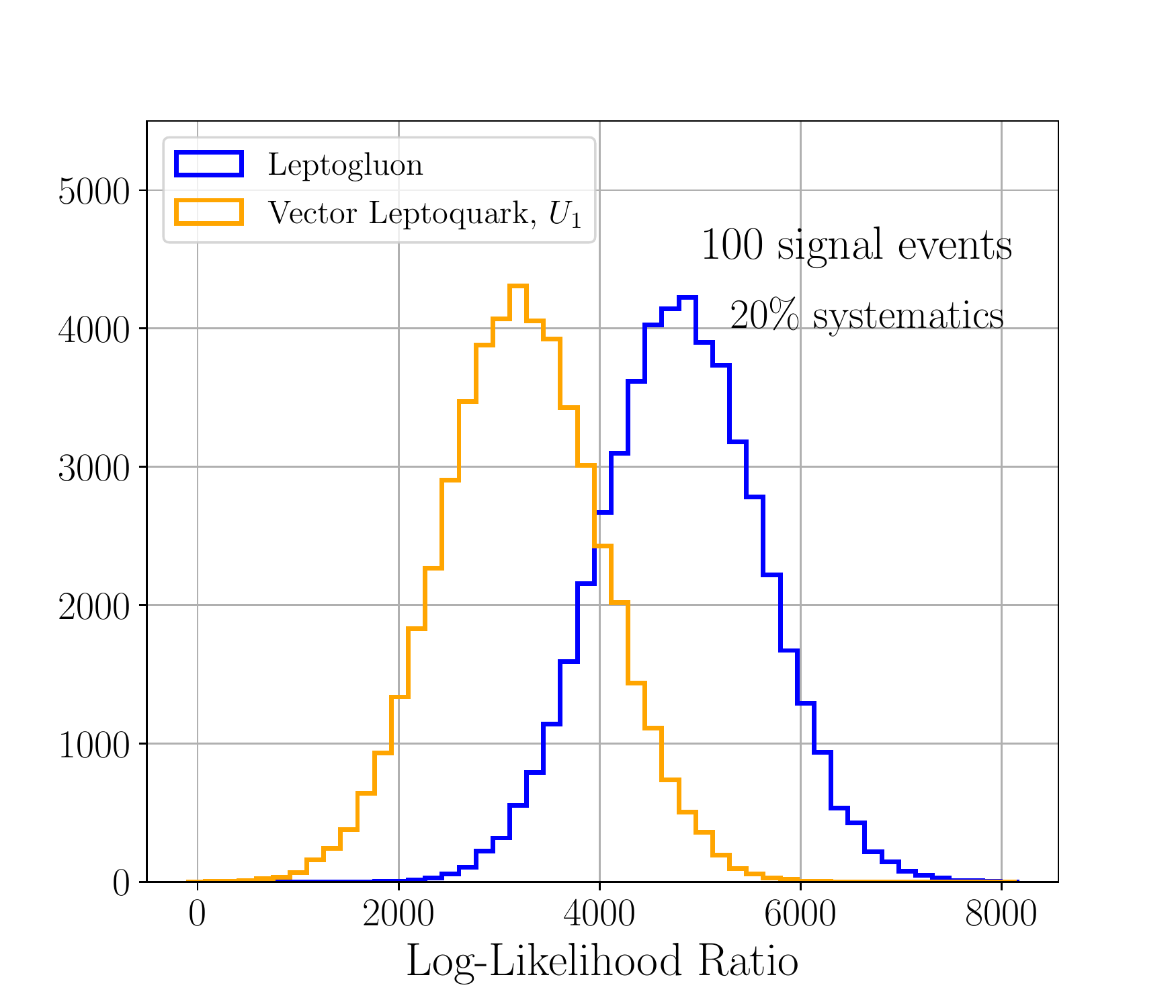}
    \caption{The distribution of the log-likelihood ratio statistic for the leptogluon {\it versus} scalar leptoquark hypothesis in the left panel, and for the leptogluon {\it versus} vector leptoquark hypothesis in the right panel. We assumed 100 signal events for each hypothesis and a 20\% systematic error in the number of events and simulated 50k pseudo-experiments.}
\label{fig:llr}
\end{figure}

\begin{figure}[!ht]
    \centering
     \includegraphics[scale=0.5]{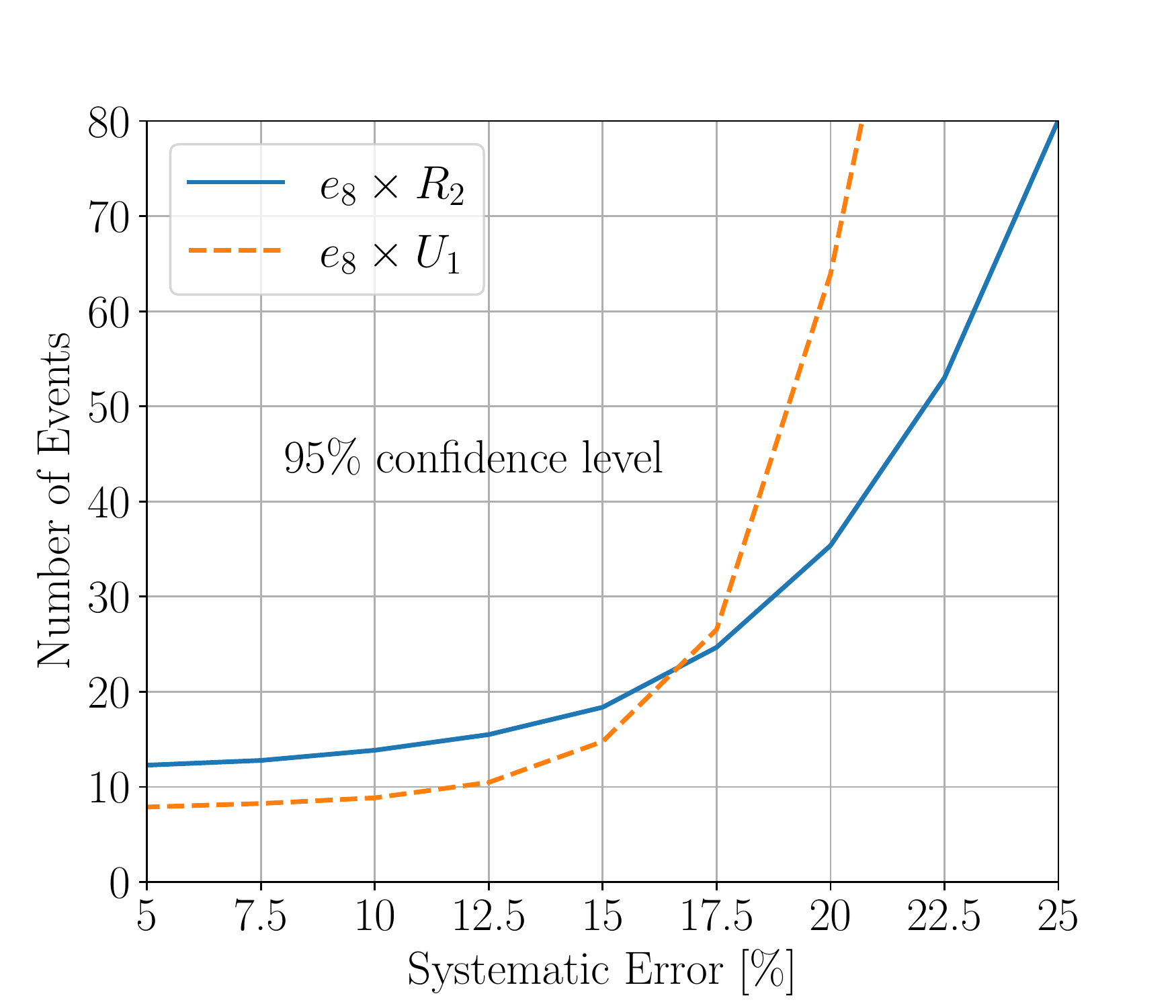}
    \caption{The number of events necessary to discern, at the 95\% confidence level, an $e_8$ leptogluon from the scalar $R_2$ (solid) and vector $U_1$ (dashed) leptoquarks  as a function of the systematic uncertainty in the number of events.}
\label{fig:nr}
\end{figure}

The log-likelihood ratio distributions corresponding to the leptogluon and the scalar and vector leptoquarks are shown in Figure~\ref{fig:llr} for a common mass of 3 TeV and assuming 100 observed events for a given hypothesis. The scalar(vector) leptoquark couples to up (down) quarks and electrons with $\lambda_{eu}=1$ and $\lambda_{ed}=0.1$, respectively, while the leptogluon with gluons and electrons occurs at a scale $\Lambda=10$ TeV, and $a_L=1,\; a_R=0$. We impose the same cuts of Eq.~\eqref{eq:cuts} except for $p_T>750$ GeV for both the leading jet and the associated lepton. We do not impose a cut on the lepton-jet invariant mass to keep the discernment power of that variable. Our results are shown in Figure~\ref{fig:nr} where we plot the  number of signal events necessary to distinguish between leptogluons and scalar and vector leptoquarks.  The number of signal events can be obtained using machine learning algorithms, for example, or by statistically subtracting the backgrounds~\cite{Pivk:2004ty,Alves:2012fb}. We also injected a systematic uncertainty on the number of events in the bins of the distributions. \smallskip

We see from Figure~\ref{fig:nr} that up to 20\% systematics, around 80 events are sufficient to tell the hypotheses apart for a 3 TeV state.
The leptogluon hypothesis might become more easily identifiable from the scalar/vector leptoquark for lower $\Lambda$ scales and heavier masses once the total width of the leptogluon resonance increases as $\sim m_{e_8}^3/\Lambda^2$ reflecting in a broader $M_{j\ell}$ distribution compared to the leptoquark case. 

We also studied the identification of the new resonance through the  asymmetry of the Collins-Soper angle distribution depicted in the central  lower panel of Figure~\ref{fig:dists}, which we defined as
\begin{equation}
    A = \frac{L\times\sigma(|\cos\theta_{CS}|>\cos\theta_{\rm cut} )-L\times\sigma(|\cos\theta_{CS}|<\cos\theta_{\rm cut})}{L\times\sigma(|\cos\theta_{CS}|>\cos\theta_{\rm cut})+L\times\sigma(|\cos\theta_{CS}|<\cos\theta_{\rm cut})}\;,
\end{equation}
where $L$ is the integrated luminosity  after all kinematic cuts and $\cos\theta_{\rm cut}$ defines the  boundaries between the central and edge regions. After computing the asymmetries, the significance of the spin hypothesis is obtained from
\begin{equation}
    Z = \frac{|A_{lg}-A_{lq}|}{\sqrt{(1-A_{lq}^2)/N}}\;,
\end{equation}
where $A_{lg}$ stands for  the leptogluon asymmetry while $A_{lq}$ is the scalar or vector leptoquark one.  $N$ is the number of events of the null hypothesis. In fact, we assumed the same number of events for both leptogluons and leptoquarks and considered leptogluons as the alternative hypothesis against the leptoquark one. \smallskip

We found that around 25 events are sufficient to tell leptogluons from vector leptoquarks, but at least 110 to tell them from scalar leptoquarks after optimizing the threshold $\cos\theta_{\rm cut}$, found to be $0.8$ in both cases. This is expected once the vector leptoquarks $\cos\theta_{CS}$ distribution is more peaked towards the edges of the distribution than the scalar ones compared to the leptogluons. The asymmetry is competitive for the vector leptoquark case compared to the log-likelihood ratio statistic when the systematic uncertainties are larger than $\simeq 15$\%. \smallskip

\section{Conclusions}
\label{sec:conclude}

 The flux of leptons from protons can be used to produce colored resonances carrying leptonic number. This is the case of scalar and vector leptoquarks, produced in lepton-quark collisions, and also the case of leptogluons produced in lepton-gluon collisions. Despite being suppressed by the initial lepton flux compared to quark/gluon fluxes, the lepton-gluon scattering permits the single production of resonances which becomes competitive to pair production, especially for heavier leptogluons. \smallskip
 
 We showed that adapting the analysis of Refs.~\cite{Buonocore:2020erb,Buonocore:2020nai}, the 95\% CL exclusion limits on leptogluons can be considerably extended compared to the most up-to-date pair production limits~\cite{Mandal:2016csb}. For example, at the 13 TeV LHC, for a 2 TeV $e_8$ leptogluon, $\Lambda/a \lesssim$ 4 (3) [2.3] TeV can be excluded at 95\% CL with 3 (1) [0.3] ab$^{-1}$ assuming a 20\% systematic uncertainty in the background rate. Considering the amount of data accumulated by the LHC  Collaborations, 139 fb$^{-1}$, $m_{e_8}\lesssim 1.8$ TeV and $\Lambda<2$ TeV, with coupling $a=1$, can be excluded by the already available LHC data. Furthermore, the 13 TeV LHC is also able to discover those leptogluons in favorable corners of the parameter space. For example, a 2 TeV $e_8$ and $\Lambda/a\sim 2.5$ TeV can be detected with $5\sigma$ significance even with 20\% systematics in the backgrounds and 3 ab$^{-1}$.\smallskip
 
 Both resonant leptogluons and leptoquarks decay into lepton-jet pairs making it difficult to tell what particle is being produced solely based on the number of events once we will not know for sure the parameters of the new physics model. It is thus important to find ways to test the leptogluon hypothesis against its leptoquark alternatives. We demonstrated that a 95\% CL distinction is possible by combining key kinematic variables in a log-likelihood ratio statistic and also by computing the asymmetry of the Collins-Soper angle distribution of the particle decays. We found that less than 100 events will suffice to tell leptogluons from scalar and vector leptoquarks if the systematic error in the signal predictions is limited by 20\% at most. The caveat is that these events should be true signals. We expect this efficient identification can be accomplished with machine learning techniques, for example. \smallskip
 
 Just like the scalar leptoquark case, studied in Ref.~\cite{Buonocore:2020erb}, we found that the search for resonant leptogluons  might benefit from lepton-gluon collisions by combining these signals with pair and associated productions initiated by quark/gluon scattering.\smallskip

 \acknowledgments
 We thank Ulrich Haisch for providing us with a modified Madgraph version to tackle initial state leptons. FSQ is supported by ICTP-SAIFR FAPESP grant 2016/01343-7, CNPq grants 303817/2018-6 and 421952/2018–0, and the Serrapilheira Foundation (grant number Serra-1912–31613).  AA is supported by CNPq (307317/2021-8).  OJPE is partially supported by CNPq grant number 305762/2019-2 and FAPESP grants 2019/04837-9 and 2022/05332-0. OJPE thanks the hospitality of the  Departament de Fisica Quantica i Astrofisica,  Universitat de Barcelona, where part of this work was carried out.

\bibliographystyle{apsrev4-1} 
\bibliography{ref}

\end{document}